# A Public Comment on NCCoE's White Paper on Privacy-Enhancing Identity Brokers

Luís T. A. N. Brandão[1,2], Nicolas Christin[2], George Danezis[3]

[1]University of Lisbon, [2]Carnegie Mellon University, [3]University College London

luis.papers@gmail.com, nicolasc@cmu.edu, g.danezis@ucl.ac.uk

## Abstract

The National Cybersecurity Center of Excellence (NCCoE) (in the United States) has published on October 19, 2015, a white paper on "privacy-enhanced identity brokers." We present here a reply to their request for public comments. We enumerate concerns whose consideration we find paramount for the design of a privacy-enhancing identity brokering solution, for identification and authentication of citizens into myriad online services, and we recommend how to incorporate them into a revised white paper. Our observations, focused on privacy, security, auditability and forensics, are mostly based on a recently published research paper (PETS 2015) about two nation-scale brokered identification systems.

## 1 Executive summary

The National Cybersecurity Center of Excellence (NCCoE)'s white paper on "privacy-enhanced identity brokers" [1] considers the development of a *privacy-preserving brokered-identification* solution (also called "building block" in the white paper). We welcome that the white paper attempts to address some privacy and security problems that have been identified (e.g., in [2]) about Connect.Gov (a governmental proposal in the US for brokering the identification / authentication of citizens into myriad online services), namely when considering "honest-but-curious" and "malicious" settings. While the white paper calls for privacy-enhancing solutions against user-attribute visibility and user impersonation by the broker, it is fundamentally incomplete by not addressing in any actionable way (e.g., goals, scenarios, solution objectives) the capabilities of the broker with respect to linkability of user-pseudonyms (a potential major enabler of undetected mass surveillance) and by completely omitting auditability and forensics aspects, whose consideration is essential for a balanced design ensuring privacy and security. In this document we express concerns about these and other aspects and provide concrete recommendations for a revision of the white paper.

---

[*]Date (before midnight CST) of original email submission to NCCoE.
[†]Simplified text, minor corrections and aesthetic adjustments.

**Remark.** This feedback applies to identity solutions based on an online central broker and with minimal user intervention (e.g., as required in Connect.Gov). Other structural solutions may be considered more adequate, e.g., based on user-aided brokering supported on appropriate privacy-enhancing technologies and cryptography.

**List of summarized recommendations.** For convenience we enumerate here a short version of each of our ten recommendations. (Sections refer to the sections of the NCCoE white paper.)

- **Five main recommendations:**

**R1.** Extend "Goal 1" (section 3) to include "unlinkability of user pseudonyms by the broker."
**R3.** In "Scenarios 2 and 3" (section 4), ask implementation of "attribute privacy" (Goal 2).
**R4.** In "Scenarios 2 and 3" (section 4), ask implementation of "unlinkability by the broker."
**R7.** In "Functional Objectives" (section 6), add a "semantic blinding" objective: a change of Y is hidden from X, for pairs (X,Y), such as (broker,user), (IDP,RP), and (RP,IDP).
**R10.** Reformulate the "foundational assumptions" (section 3), to acknowledge that a potential benefit may arise from integrating research contributions, both for development of needed solutions and for defining a threat model.

- **Five secondary recommendations:**

**R2.** Extend "Goal 4" (section 3) to require explicit "user consent" when sending attributes.
**R5.** In "Scenario 1" (section 4), encourage privacy enhancements that do not involve RPs; for example, IDP and broker can interact in a way that the broker does not see (and cannot infer) a same user pseudonym when the RP varies.
**R6.** In "Scenario 1" (section 4), encourage support (as asked by NSTIC) of transactions of types "anonymous with validated attributes" and "pseudonymous [without attributes]."
**R8.** In "Challenges" (section 5), go beyond identifying unlinkability as a challenge, by also mentioning the other main properties of "authenticity" and "attribute privacy," and identifying their simultaneous integration as challenging.

**R9.** In "Challenges" (section 5), call for description and/or definition of explicit properties of auditability and forensics and envisioned types of investigative actions they could support.

**Organization.** he remainder of this document is organized as follows. Section 2 summarizes each "concern" and presents all respective concrete recommendations. Subsequent sections elaborate further on the concerns and justification for each recommendation, in the following categories (expressions under quotes are from the white paper): privacy and security "goals" (Section 3); incremental implementation "scenarios" (Section 4); "functional objectives" (Section 5); "challenges" (Section 6); and "foundational assumptions" (Section 7).

# 2 Summarized concerns and concrete recommendations

## 2.1 Privacy and security goals

**Summarized concern 1.** The "Goal 1 (unlinkability and untraceability)" in the white paper is not addressing the linkability capabilities of the broker. See more elaborate explanation in §3.1.

**Recommendation 1 (add "unlinkability by the broker" as a Goal).** At the end of "Goal 1" (line 121 in page 9) add the sentence "*The broker cannot track or link user pseudonyms across transactions, i.e., the mediation of multiple transactions does not convey to the broker any information that could be used to infer whether the users of any two transactions are the same or a different one.*" Alternatively, use the proposed sentence as explanation of a new standalone goal, e.g., "Goal 5" or "Goal 1b" called "unlinkability by the broker," and (based on [2]) rename "Goal 1 (Untraceability and unlinkability)" to "Goal 1a (edge unlinkability)," where edge refers to the IDPs and RPs.

**Summarized concern 2.** Goal 4, referring to *when* user attributes should be transmitted, does not mention the need for "user consent," but this aspect should be explicitly considered at design time. See more elaborate explanation in §3.2.

**Recommendation 2 (add "user consent" to Goal 4).** At the end of the title of Goal 4 (line 139, page 9), add the following expression: "*and consented by the user*." Then, in line 141, between the first and second sentence of the explanation of Goal 4, add the following sentence: "*Also, attributes should only be transmitted between an IDP and the broker if and after the user gives informed consent about transmission of said attributes.*"

## 2.2 Incremental implementation scenarios (1,2,3)

**Summarized concern 3.** The incremental implementation "scenarios" (1, 2, 3) in the white paper are not calling for achievement of "attribute privacy" ("Goal 2" identified in the white paper), but they should. See more elaborate explanation in §4.1.

**Recommendation 3 (require "attribute privacy" in Scenarios 2 & 3).**

a) In the titles of scenario 2 and 3 (lines 284 and 295, respectively), replace "attribute delivery" by "*attribute delivery and **privacy***."

b) In scenario 2, in line 285, replace "Goal 1 and Goal 4" by "Goal 1, **Goal 2** and Goal 4."

c) Copy or move (with minor adaptations) to the body of the explanation of "scenario 2," immediately after line 294 in section 4, the sentence from lines 113–116 (in the preamble of paragraph "Goals" in section 3): "*Specifically, this building block seeks innovative ways to encrypt the attributes of a logged-in user, such that the identity broker can never decrypt the attributes and gain access to personal information – while retaining an architecture in which RPs and IdPs do not know each other's organizational identities – i.e., double-blind.*"

d) In scenario 3, add a new item to the list (line 304) of actions foreseen by a malicious broker (immediately after line 308): "*4. Attempts to perform a man-in-the-middle attack to obtain the encryption keys that would enable decryption of user attributes intended for the RP, and/or obtain user pseudonyms intended solely for the RP or the IDP.*"

**Summarized concern 4.** The incremental "implementation scenarios" are also not calling for achievement of "unlinkability of user-pseudonyms by the broker," but they should. See more elaborate explanation in §4.2.

**Recommendation 4 (require "unlinkability by broker" in Scenarios 2 & 3).**

a) Change the "authentication and attribute delivery..." titles of scenarios 2 and 3 (lines 284 and 295) to "*authentication, **pseudonym unlinkability**, and attribute delivery and privacy...*"

b) In the explanation of "scenario 2" (and consistently with our Recommendation 1 and aligned with Recommendation 3), add the following text after line 294: *Specifically, this building block seeks innovative ways to prevent the broker from seeing persistent user-pseudonyms in transit between the IDP and the RP, including prevented from inferring any user-pseudonym linkable to other transactions, while also ensuring that user-pseudonyms are transformed* *appropriately for each RP, and while retaining an architecture in which RPs and IdPs do not know each other's organizational identities.*

**Summarized concern 5.** Scenario 1, representing the baseline migration of a RP from a setting of multiple IDPs to a setting with a single broker, should be more flexible to allow privacy enhancements that only involve the IDP and broker. See more elaborate explanation in §4.3.

**Recommendation 5 (in Scenario 1 encourage IDPs to enable "weak unlinkability by the broker").** In scenario 1, replace lines 282–283 by the following phrase and append it to the end of line 281: "*In summary, with respect to RPs this baseline scenario considers that the migration to a brokered identification system corresponds simply to having the RP integrating with the broker as being a new IDP.*" Then, add the following new paragraph: "*Since IDPs have a privileged flexibility, aligned with their incentive to be certified and compliant with a brokered identification system, in this scenario the relationship between the IDP and the broker can be augmented to enable enhancements to user privacy. For example, without any protocol change at the RPs, the IDP and the broker can interact in a way that the broker does not see (and cannot infer) user pseudonyms that remain equal (or are linkable) when the RP changes.*"

**Summarized concern 6.** The current focus is on transactions of the type "pseudonymous with attributes" (often uniquely identifiable), but NSTIC [3] calls for enabling "a variety of transactions," e.g., including "anonymous with validated attributes," and "pseudonymous [without attributes]." See more elaborate explanation in §4.3.

**Recommendation 6 (in Scenario 1 encourage a design with flexibility for a variety of types of transaction).** In the end of Scenario 1, immediately before line 284, add the following paragraph: "*Also starting with this baseline scenario, it is important that the system design maintains a flexibility for 'a variety of transactions', as asked by NSTIC, including of types 'anonymous with validated attributes' and 'pseudonymous without attributes'. For instance, an implementation should support transactions where attributes are validated (e.g., an age range), but neither the broker nor the RP receive from the IDP (nor are in a position to infer from the elements of the transaction) any user pseudonym that would persist beyond said transaction. As another example, the system should also support transactions where the RP and the broker are able to signal to the broker and the IDP, respectively, that no attribute or attribute validation is required for a transaction.*"

## 2.3 Functional objectives

**Summarized concern 7.** The "triple blinding" functional objective, i.e., that "entity X does not know identity of entity Y," for some pairs (X,Y), should be extended to include *semantic blinding*, i.e., that "change of Y is hidden from X" across any two transactions, for some pairs (X,Y), most importantly for the pair (user,broker). See more elaborate explanation in §5.

**Recommendation 7 (add "semantic blinding" as a Functional Objective).** Augment "Table 2 – Function Objectives," renaming the third row from "Triple blinding" to "*Triple blinding per transaction*" and then add a new row labeled "*Semantic triple blinding across transactions*," containing the following content: "*Across any two transactions mediated by the broker:*
  a. *a 'change of RP is hidden from IDP' (i.e., for two transactions involving the same IDP, the IDP does not learn whether the RP is the same or has changed);*
  b. *a 'change of IDP is hidden from RP' (i.e., for two transactions involving the same RP, the RP does not learn whether the IDP is the same or has changed);*
  c. *'the broker cannot link users across transactions' (i.e., it cannot tell whether users in any two transactions are the same or different, regardless of the involved RPs and IDPs).*"

## 2.4 Challenges

**Summarized concern 8.** The white paper (section 5) positions unlinkability by the broker as a "challenge" and not as a "goal." Conversely, the white paper defined "attribute privacy" and "authenticity" (i.e., resilience against impersonation) as "goals," but in Section 5 does not mention them as "challenges." In practice these later two also have inherent challenges (e.g., not achieved in Connect.Gov). We find more value in considering all three main properties as goals with inherent challenges, and in explicitly considering as a main challenge the implementation of all those properties in an integrated way. See more elaborate explanation in §6.1.

**Recommendation 8 (discuss challenges consistently across essential properties, and identify new challenges).** In section 5, instead of singling out *unlinkability by the broker* as a conflicting requirement, make adjustments to highlight as main challenge the simultaneous achievement of "user authenticity, attribute privacy and unlinkability by the broker," as follows:

  a) Replace the third paragraph (page 12, lines 333–228) with the following: "*In spite of possible standard stand-alone solutions for some isolated properties of privacy and security (e.g., authentication, attribute privacy, unlinkability by the broker), it is more challenging to ensure (and define) an integrated solution with all properties. For example, the broker knows who is* *the IDP and RP, so that it is able to mediate the user in the authentication protocol, but it does not need, and should not learn, information about the user, which means is should not see the information flowing between IDP and RP. Simultaneously, information that reaches each RP may have to be transformed while in transit from the IDP, e.g., user pseudonyms have to be changed to ensure that different RPs receive different user pseudonyms. The challenge of integrating these and other properties (e.g., including auditability and selective forensic investigations) should be considered explicitly at a design phase, aligned with NSTIC's requirements, e.g., as derived by NSTIC-NPO [4], that "organizations" (including the broker and IDPs) shall "minimize data aggregation and linkages across transactions, [...] use privacy-enhancing technology that: minimizes the transmission of unnecessary information; [...] minimizes the ability to link credential use among multiple service providers."*

b) Between the third and fourth sentence (line 343) of the fourth paragraph, talking about existing cryptographic solutions vs. existing market products, insert the following sentence: *"Thus, it is also important to tackle the challenges associated with bringing to implementation and/or commercial viability those researched solutions that are likely to enable a better integration between all desired properties."*

**Minor note:** In order for section 5 to be more focused on identified "challenges," it is possible to: move its first paragraph (lines 320–324), related to business value, to section "2 (Business value)" after line 64; move its second paragraph (lines 325–332), related to NSTIC vision, to the preamble of the paragraph "Goals" in section 3 (between lines 112 and 113).

**Summarized concern 9.** Auditability and forensic actions are ignored in the white paper, but their consideration is essential for a design that enables their balanced integration with privacy and security. See more elaborate explanation in §6.2.

**Recommendation 9 (consider auditability and forensics).** In the end of section 5, after mentioning (in lines 345–346) the goal of illustrating "how to *simultaneously* satisfy" "accountability" and other privacy and security properties, insert the following paragraph: *"Specifically, it is important to identify and describe at design time the main envisioned actions of auditing and selective forensic investigation that may contribute to accountability, so that the system design may achieve a balanced integration with properties of privacy and security, and also be aligned with transparency to and adequate expectations by the users. For example, the privacy and security analysis of the system should take in consideration the case of collusion (e.g., based on a subpoena) between a broker and an IDP and/or a RP compelled to help, or of a mediated auditing using information from several parties. It is paramount to consider how to enable selective forensic capabilities that*

*(even in an adversarial case) do not jeopardize the privacy and security of transactions of users and/or related to RPs that are not of the interest of a legally allowed investigation."*

## 2.5 Foundational Assumptions

**Summarized concern 10.**

- "Foundational assumption 1" (lines 230–232), that current market's resources are able to satisfy the goals in the white paper, is in practice limiting the goals set forth in the white paper, in a way that contradicts the guiding principles of NSTIC. The goals needed for an actual privacy-enhancing solution currently require an alignment with NSTIC's "seek" "to encourage new solutions where none exist" and NSTIC's goal (4.1) to "Drive innovation through aggressive science and technology (S&T) and research and development (R&D) efforts." The assumption is also opposed to several acknowledgments in the white paper, about the need for innovative solutions. Thus, the assumption needs to be revised.

- "Foundational assumption 3" (lines 241–244) puts in potential conflict the benefits of existing solutions vs. the benefits of solutions that still need to arise from directed research. A more collaborative approach is possible, by simultaneously praising the potential contribution from solutions existing in the market and encouraging complementary research that may provide missing or better components.

See more elaborate explanation in §7.

**Recommendation 10 (in the Foundational Assumptions, frame the role of research in a positive synergistic way).** (For simplicity, our suggestion follows the format of the current white paper in that the "assumptions" are not only assumptions but also a statement of goals.)

a) Reformulate and merge the foundational assumptions 1 (lines 230–232) and 3 (lines 241–244), as follows: "*The technologies, algorithms, standards and processes already existing in today's market have **the potential for significant contribution** to satisfy the goals of this building block. Whenever adequate to fulfill the security and privacy-enhancement requirements of the building block, these market capabilities should be used, if necessary including changes and augmentations to existing protocols and profiles. Complementary, the development of the building block should take advantage of **improvements that may arise from concurrent research efforts**, and use them whenever demonstrated that their effectiveness and/or efficiency may surpass the benefit of market dominant protocols and profiles.*"

b) Add the following complementary "assumption": "*For the development of the intended building block, the state-of-the-art knowledge and practice in security engineering and cryptography research have the potential for significant contribution to the definition and formalization of an explicit threat model, where adversarial goals and capabilities are considered for all inherent parties of the protocol, including malicious collusion between parties, and eventual external adversaries. The development of the building block must take in consideration an adequate threat model that considers the possibility that parties may be maliciously compromised with the intent to disrupt the privacy and security goals of the system.*"

## 2.6 Typo

In the "disclaimer" of the white paper (page 3), the expression "Such identification does represent an exhaustive list of commercially available technologies" probably intends to be "*Such identification does **not** represent an exhaustive list of commercially available technologies.*"

# 3 Privacy and security goals

In section 3 ("Description"), the white paper enumerates and explains four goals for the *Privacy-Enhanced Identity Brokers* "building block":

- Goal 1. Untraceability and unlinkability

- Goal 2. The identity broker cannot access user attributes.

- Goal 3. A compromised or malicious broker cannot impersonate a user.

- Goal 4. User attributes are only provided when requested by the RP.

## 3.1 Unlinkability by the broker

**Concern 1 (unlinkability goal).** Unfortunately, the enumerated "Goals" do not state as desirable the property of "user-pseudonym unlinkability by the broker" – one whose lacking is most suitable to enable undetected and unaccountable mass surveillance by a possibly compromised broker. Specifically, the detailed description of "Goal 1 – untraceability and unlinkability" (page 9, line 119) only considers unlinkability between the IDP and the RP, ignoring what the broker is able to infer. Conversely, the NSTIC-NPO "derived requirements" [4, Req. 5] says that "Organizations shall use privacy-enhancing technology that [...] minimizes the ability to link credential use among multiple service providers." Also, NIST [5] asks that "unlinkability assures that two or more related events in an information processing system cannot be related to each other." Thus, this property should be explicitly considered for any privacy-enhanced identity broker. While brokering prevents an IDP from tracking users across RPs, so should the broker be prevented from tracking users (including user pseudonyms) across different RPs.

Related to this property, the white-paper acknowledges in its abstract that: "the broker, if *deployed incorrectly*, is in a significant position of power, as it creates the potential to track or profile an individual's transactions." Unfortunately, the important counter-measure of ensuring *user-pseudonym unlinkability by the broker* is not effectively represented in the actionable paragraphs of "goals" (section 3), "assumptions" (section 3), "scenarios" (section 4) and "functional objectives" (section 6). By leaving out this essential property the white paper is thus promoting that brokered-identification systems be "incorrectly deployed" (to reuse the expression from the abstract of the white paper). See concrete recommendation in §2 (Recommendation 1).

**Remark (linkability of user-pseudonyms by the broker also degrades attribute privacy).**
Linkability of user-pseudonyms is worrisome on its own. Furthermore, in an adversarial model it introduces a nefarious consequence of degrading attribute privacy, in the following sense. Consider that a solution is in place to ensure attribute privacy (i.e., that even a malicious broker is not able to see user attributes), but that the system allows the broker to always see, for each user, the same user-pseudonym from the IDP – this is the case in Connect.Gov. Then, a single collusion with a RP (or a data-breach with public leakage of the private data of a single RP) gives to a honest-but-curious broker enough information to correlate user attributes to transactions of users across all RPs. In other words, regardless of how "strongly secure" a certain $RP_1$ is, a single intrusion into some "weak" $RP_2$ enables the broker to infer the attributes of the respective users that have interacted with $RP_1$ via the brokered identification system.

## 3.2 User consent for attribute disclosure

**Concern 2 (lack of *user consent* in Goal 4).** Since Goal 4 is about determining *when* attributes should be transmitted, the aspect of *user consent* should be explicitly included. This is relevant from a design standpoint because it may affect the transaction protocol, involving an action from the user, still at the stage of even allowing the RP to request a set of attributes. This may also better promote a brokered identification solution that explicitly prevents an adversarial broker from forging the request of attributes from RP to IDP. (We notice that "user consent" has been mentioned in the text of Goal 2, when describing the flow of information, but there the emphasis (including in the title) is about the broker not seeing attributes between the IDP and the RP.) See concrete recommendation in §2 (Recommendation 2).

# 4 Incremental implementation scenarios (1,2,3)

In section 4 ("Scenarios"), the white paper starts by defining a *baseline* ("scenario 1") of migration from a solution of multiple IDPs to a solution with a single-broker, without any privacy and security considerations with respect to adversarial capabilities of a possibly compromised broker. Then, the white paper asks for resilience, with respect to "*authentication* and *attribute delivery*," against an adversarial broker, in two incremental scenarios:

- Scenario 2: resilience against an honest-but-curious broker

- Scenario 3: resilience against a malicious broker

## 4.1 Attribute privacy

**Concern 3 (missing "attribute privacy").** In lines 264–265, the white paper claims that "The following [three] scenarios establish incremental capabilities to achieve the goals of this white paper." For example, the title of scenario 2 alludes to "attribute delivery," (which relates to "Goal 4") and its explanation starts by saying that "In Scenario 2, Goal 1 and Goal 4 are achieved." However, unfortunately, Goal 2 ("The identity broker cannot access user attributes") is ignored in the description of scenarios 2 and 3, even though it is immediately relevant in the case of a honest-but-curious broker. Leaving out "Goal 2" from a plan of incremental implementation scenarios means effectively ignoring the capabilities of an honest-but-curious broker and with more nefarious consequences in case of a malicious broker. We also notice that "attribute privacy" was already a requirement of the original FCCX solicitation [6], which was also not respected in the subsequent Connect.Gov implementation (e.g., see Q&A 2.3 in [7]).

Given the current confusion in the white paper (i.e., forgetting to mention the *attribute privacy* goal in the incremental scenarios, instead alluding only to the operational aspect of when to deliver attributes), we suggest that in the titles of scenarios 2 & 3 the aspect of "attribute privacy" is explicitly stated, and that "Goal 2" is explicitly mentioned in scenario 2, and that scenario 3 includes an explicit reference to attribute privacy in case of a malicious broker.

See concrete suggestion in §2 (Recommendation 3).

**Remark.** We notice that attribute privacy is later included in Section 6 ("Desired Solution Objectives") as a "capability" example of the "confidentiality" security objective. Yet, we emphasize that explicitly mentioning "attribute privacy" in the description (and titles) of scenarios 2 & 3 would more adequately promote that the property be actually implemented in the "scenarios."

## 4.2 Unlinkability of user-pseudonyms

**Concern 4 (unlinkability of user-pseudonyms).** Unfortunately, in the white paper neither scenario 2 nor 3 encompass the desirable (e.g., desirable by NSTIC) privacy property of "unlinkability (by the broker) of user-pseudonyms" [2]. However, a privacy-enhancing brokering solution should ensure that across any two identification transactions the broker does not learn (from values inherently exchanged in the authentication protocol) whether the interacting user is the same or has changed. This property can also be considered in a progression:

1. **Weak unlinkability (by the broker) of user-pseudonyms across RPs.** Across any two identification/authentication transactions with different RPs, an adversarial broker – honest-but-curious (scenario 2) or malicious (scenario 3) – must not be able to infer whether the same user or different users were involved. This means that the broker must not see a user identifier (e.g., pseudonym) that remains persistent across transactions involving the same user but different RPs. At the same time, the IDP (who knows that the authenticating user is the same across two transactions), must neither know who is the RP in each transaction, nor know if the RP changes or remains the same. As mentioned, this exact property suits what has been identified by NSTIC-NPO as a NSTIC requirement: "Organizations shall use privacy-enhancing technology that [...] minimizes the ability to link credential use among multiple service providers."

2. **Strong unlinkability (by the broker) of user-pseudonyms.** As a natural strengthening of "weak unlinkability," "strong unlinkability" prevents the broker from tracking users / user-pseudonyms across any two transactions, regardless of whether the IDP and RP remain the same or change. For such property, the hub should not see any user persistent identifier (e.g., pseudonym), not even the one that the RP requires to map the transaction into a local user account. Also this property can be considered in the honest-but-curious (scenario 2) and in the malicious (scenario 3) adversarial models.

See concrete suggestion in §2 (Recommendation 4).

## 4.3 More flexible scenario 1, toward better privacy

We notice that even the Connect.Gov implementation, which corresponds roughly to an approximation of scenario 1 (in that it does neither provide user authenticity under a malicious model, nor provide attribute privacy under honest-but-curious or malicious models), already enables some forms of unlinkability, e.g., between the IDP and RP, and between any two RPs. There, the broker changes the user pseudonym received from the IDP (also known as MBUN – "meaningless but unique number") into a new one for each RP (also known as rPAI – persistent anonymous identifier for the RP), thus preventing colluding RPs from correlating user pseudonyms. The pseudonym transformation by the broker, in Connect.Gov, fits scenario 1 because it does not imply any change to the RP, i.e., in the perspective of the RP the broker is just like another IDP. In contrast, IDPs are entities that must be flexible to adapt to the requirements of the broker, which may for example include being FICAM certified. Thus, in the same way that even in Connect.Gov the broker is not simply a relay between the IDP and the RP, so should an initial implementation (i.e., scenario 1) arising from the call of this white paper take some privacy considerations, namely whenever those do not involve any adjustment to the RPs. Below we discuss two concerns that could be addressed (even if just optionally per IDP) immediately in scenario 1.

**Concern 5 (a closer look at scenario 1).** The description of "Scenario 1" has two contradictory statements: in line 269 it says that "It achieves previously specified Goal 1 (untraceability and unlinkability)," but then in line 276 it says "The baseline scenario does not accomplish any of the privacy goals desired herein." Indeed, it does (e.g., Connect.Gov) achieve "goal 1 – unlinkability and untraceability" (in the limited sense defined in the white paper), which is privacy related, with respect to honest-but-curious IDPs and RPs. (However, it does not achieve Goal 1 in the augmented sense that we propose in Recommendation 1.) We argue that even Scenario 1 should already contemplate and encourage (e.g., optional per IDP) some privacy goals – those for which the integration between RP and broker remains as a usual integration between RP and a new IDP. This is possible by recognizing the privileged flexibility of IDPs, aligned with their incentive to be certified and compliant with a brokered identification system.

For example, the property of "weak unlinkability (by the broker) of user-pseudonyms across RPs" [2] can be based on an adaptation of the interaction between IDPs and the broker (i.e., not involving the RPs). The property ensures that for any user the respective pseudonyms seen and received by the broker varies with the RP, while at the same time preventing the IDP from knowing who is the RP or if it has changed. In other words, this means eliminating the "MBUN" (meaningless but unique number) existing in the Connect.Gov system. Optimistically, some IDPs may find motivation to accommodate this property from the start, e.g., to be able to advertise a distinctive privacy-enhancing capability. This also provides a first step toward *strong unlinkability*, which may then be achieved in scenarios 2 and 3. See concrete suggestion in §2 (Recommendation 5).

**Concern 6 (need for explicit support of anonymous and pseudonymous transactions).** NSTIC envisions an Identity Ecosystem that "will enable a variety of transactions, including anonymous, anonymous with validated attributes, pseudonymous, and uniquely identified." However, the current development (e.g., Connect.Gov) seems mostly focused on the later type, i.e., "pseudonymous with [often uniquely identifiable] attributes." Thus, we argue that the white paper should more explicitly call for flexibility to handle transactions of other types, e.g., "anonymous with validated attributes" and "pseudonymous without attributes." For example, in an anonymous transaction with validated attributes, the IDP should not send a persistent user pseudonym to the hub, but rather just validate a predicate of user attributes. This support suits scenario 1 because it can be achieved without changes to the way RPs integrate with the broker. Also, enabling this support from the beginning (i.e., since scenario 1), sets a better basis for privacy and security considerations in scenarios 2 and 3. See concrete suggestion in §2 (Recommendation 6).

# 5   Functional objectives

**Concern 7 (semantic blinding).**   In section 6, Table 2 (page 17), the white paper defines the concept of "triple blinding," with three properties of the form "entity X does not know identity of entity Y" [in a transaction], where the pairs (X,Y) are (IDP,RP), (RP,IDP) and (broker,user). We denote these as *static blinding* properties. For example, *static blinding* may be satisfied in a hub-based brokered identification system if "attribute privacy" is ensured and if the hub effectively shields the identities of the IDP and RP from one another. However, a major concern is that the white paper is missing the concept of *semantic blinding*, which would imply that a "change of Y is hidden from X" across any two transactions. Such property is particularly important for the pair (broker,user), i.e., the broker should not be able to correlate different transactions to the same user (regardless of attributes or pseudonyms required by the RP). This is what we call "unlinkability of user-pseudonyms by the broker".

Semantic (broker,user) blinding is most essential in the scope of a privacy-enhancing system; semantic (IDP,RP) blinding is somewhat trivial to achieve, and is even satisfied in Connect.Gov; in terms of user privacy, semantic (RP,IDP) blinding is arguably less important. All these three pairs have been considered in [2]. See concrete suggestion in §2 (Recommendation 7).

# 6   Challenges

Section 5 of the white paper is about "current building block challenges."

## 6.1   Challenges associated with goals

**Concern 8 (challenges inherent to goals).**   In section 5, the white paper states "unlinkability [by the broker]" as a *challenge* (one which has not been identified as a goal). In contrast, the white paper states "attribute privacy" and "authenticity" as *goals* (section 3), but in section 5 it does not include explicit considerations about them, even though they are still unresolved challenges. For example, even Connect.Gov (the current public attempt to develop an actual nation-scale brokered identification system for citizens in the US) did not achieve a solution that guarantees attribute privacy or authenticity in an adversarial model. We propose that the three mentioned properties – authenticity, attribute privacy and unlinkability by the broker – be posed in equal standing with respect to being a current challenge and a goal.

Also in Section "5 (current building block challenges)," line 327, the white paper acknowledges NSTIC's vision of an Identity Solution where unlinkability is indeed satisfied. However, the white paper then follows to say that such kind of unlinkability (line 337) is a *"conflicting requirement"* when considered versus (line 334) the need of "information about all the entities involved in a particular transaction to that it can help guarantee the integrity and confidentiality of the transaction, as well as the information contained within the transaction." We notice here that it is **technically incorrect** to say (third paragraph, lines 333–228) that the broker needs to know "the information contained within the transaction in order to help guarantee its integrity or confidentiality." For example, in a technically possible privacy-enhancing brokered identification solution, the broker **does not need** to see or known identity attributes or pseudonyms of the user in order to mediate the user across a transaction between an IDP and RP. And indeed, in a privacy-enhancing solution the broker **should not** learn such information.

In summary, we think that it would be more valuable that Section 5 in the white paper identifies that several identified goals have inherent challenges, and that an integrated achievement of all goals is on its own a challenge to consider. See concrete suggestion in §2 (Recommendation 8).

## 6.2 Auditability and forensics

**Concern 9 (auditability and forensics).** The white paper does not provide any explicit note about auditability and forensics, which may be used to promote accountability. However, both from NSTIC requirements and initial FCCX requirements [6] we know that auditability and forensic actions are foreseen, namely when required by law. Thus, we suggest that envisioned auditability and forensics use-cases be made explicit ([2]).

There is a privacy and security danger of not explicitly defining requirements of auditability and forensics. An unprepared system may have its privacy and security globally broken when breached (even legally) by one such event of mandated auditability or forensic action, even if such action only has a local intention (e.g., single user, or transactions with a single RP). If instead these actions are considered at a design stage, then it is possible to ponder in advance how to integrate them in a privacy-preserving system. Auditability of the broker may be pondered in a way that respects the privacy of the users involved in the respective transactions. A property of *selective* *forensic disclosure* [2] may allow selective investigation of user transactions, i.e., selective per user and per RP or IDP, without enabling total undetected disclosure by default (all transactions of all users across all RPs and IDPs – i.e., undetected mass surveillance). In a possible solution, the broker is only selectively allowed to break pseudonym unlinkability or attribute privacy properties, if and when receiving compelled help (e.g., upon a subpoena) from the respective IDP and/or RPs. This then allows each RP and/or IDP to reveal, e.g., at certain intervals of time, how many compelled collaborations they participated in, and how many users were investigated. It also possible to define different levels of forensic capabilities, such as "coarse-grained" vs. "fine-grained", e.g., respectively not restricted vs. restricted to a single RP. Even if it is arguable the need and/or the way in which these technical capabilities are integrated with other privacy and security properties, their explicit definition is paramount at a design phase, to ensure promotion of the principle of *transparency* to users. See concrete recommendation in §2 (Recommendation 9).

# 7 Foundational assumptions

In section 3, paragraph "Assumptions" (page 8, line 227), the white paper states three "foundational assumptions." The first and the third "assumptions" (which besides assumptions also state goals/objectives) affect the paths toward devising a robust identity solution.

- Assumption 1. "The technologies, algorithms, standards, and processes already exist in today's market, and are available to fully satisfy the goals of this building block; the objective is to utilize state of the market capabilities."

- Assumption 3. "The goal of this building block is to consider how to augment existing, market dominant protocols; it is **not** to develop or research new protocols. However, we recognize that changes to existing protocols and profiles may be necessary to fulfill the building block's privacy enhancement requirements."

**Concern 10 (Inadequacy of foundational assumptions 1 and 3).** In contrast to "Assumption 1," the white paper acknowledges (lines 339–340) that "the current standards and product market does not have non-proprietary mechanisms to employ a privacy-enhancing solution in identity brokers." Thus, as a logical inference from Assumption 1 and the referred sentence, one would conclude that the building block being promoted by the white paper is a brokered-identification system that is **not** privacy-preserving. We thus think that Assumption 1 may lead to an inadequate pressure to compromise the "privacy-enhancing" guiding principle of NSTIC.

Assumption 1 (that today's Market is enough) is misaligned with the developmental context of a privacy-enhancing brokered identification system. For example, the preamble of the white paper (page 2, before the abstract) states that "NCCoE building blocks address technology gaps." Also, the properties of "authenticity" and "attribute privacy" requested by the white paper (as Goals to be achieved in adversarial models) have not yet been achieved (e.g., in Connect.Gov) and thus require development, e.g., the white paper specifically calls (lines 113–115) for "innovative ways to encrypt the attributes of a logged in user." Also, NSTIC "seeks to promote the existing marketplace, encourage new solutions where none exist," and asks that "the government should not over-define or over-regulate the existing and growing market for identity and authentication services." Aligned with this, it is valuable to be open to accept developments (e.g., from research) that may contribute with new solutions to the market of identity and authentication services.

A main danger of the current assumptions is that it may lead designers and developers in this context to chose considering only problems for which a tailored solution already exists, ignoring other essential privacy and security problems that have already been identified and whose solution is possible through a developmental effort. Specifically, we argue that the white paper incurs exactly this kind of choice – for example, even though there is awareness that linkability of user-pseudonyms may enable undetected mass surveillance by a compromised broker [2], the white paper does not include said unlinkability as a goal.

The above concern is further consolidated by the current form of assumption 3. We welcome that the white paper considers the idea of "augmenting [...] existing protocols and profiles," and we concede to preferring existing solutions when they are adequate. However, we are concerned that in something denoted as a "foundational assumption" a *goal* is expressed in the negative: "*The goal: [...] is **not** to develop or research new protocols*." We think that there is a danger of this statement being present in a "foundational assumption," in the sense that it may subvert the direction of an enumeration of privacy and security goals (e.g., unlinkability, attribute privacy, authentication) needed in a privacy-enhancing brokered identification system.

As already discussed, some of the needed properties (including some of those identified as goals) have not yet been achieved in practice (i.e., and in an integrated way), e.g., "attribute privacy," in case of a malicious broker, Thus, they need to be developed, lest the privacy-preserving guiding principle of NSTIC be compromised. Thus, we recommend a reformulation of the assumptions in the white paper, which can still state a preference for solutions already existing in the market, and highlighting their potential benefit, while also explicitly welcoming new research results, which may turn out to provide, more timely, important contributions toward an actual privacy-preserving brokered identification system. This makes even more sense considering that the white paper also acknowledges that "Research exists that identify cryptographic solutions to meet the goals outlined in [the white paper]." Furthermore, It is useful here to recall Goal 4.1 of A Public Comment on NCCoE's White Paper on Privacy-Enhancing Identity BrokersNSTIC [3]: "Drive innovation through aggressive science and technology (S&T) and research and development (R&D) efforts."

The overall concern about the Assumptions extends also to effects on the still-needed endeavor of explicitly formalizing privacy, security and other (e.g., auditability and forensics) requirements in integration with a threat model (e.g., as called for in Section 4 ("Scenarios") in the case of honest-but-curious (line 291) and malicious (line 209) models). We are concerned that a limitation to market products could potentially induce designers / developers to only consider adversarial models that may be mitigated with existing market products. Such approach could render the development oblivious to real not-addressed vulnerabilities. Instead, we propose a reformulation that recognizes that an adequate solution (whatever it should be) under an adequate threat model (whatever it should be) may receive contributions both from market products and from research (without prejudice of stating a preference for already existing elements), both for development of solutions and development of a threat model. See concrete recommendation in §2 (Recommendation 10).